\documentclass[12pt]{article}

\usepackage[preprint]{neurips_2023}

\usepackage[utf8]{inputenc} 
\usepackage[T1]{fontenc}    
\usepackage{booktabs}       
\usepackage{amsfonts}       
\usepackage{nicefrac}       
\usepackage{microtype}      
\usepackage{xcolor}         
\usepackage[title]{appendix}

\usepackage{wrapfig}
\usepackage{paralist}  
\usepackage{enumitem}

\usepackage{algorithmic}
\usepackage{algorithm}
\usepackage{caption}
\usepackage{subcaption}
\usepackage{graphicx}
\usepackage{natbib}
\usepackage{amsmath}
\usepackage{amssymb}
\usepackage{mathtools}
\usepackage{amsthm}
\usepackage{dsfont}
\usepackage{subfiles}
\usepackage{tikz}
\usepackage{bbm}
\usepackage{comment}

\theoremstyle{plain}
\newtheorem{theorem}{Theorem}

\newtheorem{corollary}{Corollary}
\theoremstyle{definition}

\newtheorem{assumption}{Assumption}
\theoremstyle{remark}

\newenvironment{customthm}[1]{\renewcommand\thetheorem{#1}\theorem}{\endtheorem}
\newenvironment{customcorr}[1]{\renewcommand\thecorollary{#1}\corollary}{\endcorollary}

\newcommand{\Pa}[2]{\textit{Pa}_{#2}(#1)}

\newcommand{\D}[0]{\mathbf{D}}
\newcommand{\V}[0]{\mathbf{V}}

\newcommand{\W}[0]{\mathbf{W}}
\newcommand{\U}[0]{\mathbf{U}}
\newcommand{\X}[0]{\mathbf{X}}

\newcommand{\Y}[0]{\mathbf{Y}}
\newcommand{\Z}[0]{\mathbf{Z}}
\renewcommand{\O}[0]{\mathbf{O}}

\newcommand{\E}[0]{\mathbb{E}}

\newcommand{\x}[0]{\mathbf{x}}

\newcommand{\G}[0]{\mathcal{G}}
\newcommand{\M}[0]{\mathcal{M}}

\newcommand{\Cov}[2]{\mathrm{Cov}(#1, #2)}
\newcommand{\Var}[1]{\mathrm{Var}(#1)}

\title{A Cross-Moment Approach for Causal Effect Estimation}

\author{%
  Yaroslav Kivva\\
  School of Computer and Communication Sciences\\
  EPFL, Lausanne, Switzerland \\
  \texttt{yaroslav.kivva@epfl.ch} \\
  \And
  Saber Salehkaleybar\\
  School of Computer and Communication Sciences\\
  EPFL, Lausanne, Switzerland \\
  \texttt{saber.salehkaleybar@epfl.ch} \\
  \AND
  Negar Kiyavash\\
  College of Management of Technology\\
  EPFL, Lausanne, Switzerland \\
  \texttt{negar.kiyavash@epfl.ch} \\
}

\begin{document}

\maketitle

\begin{abstract}
 We consider the problem of estimating the causal effect of a treatment on an outcome in  linear structural causal models (SCM) with latent confounders when we have access to a single proxy variable.
Several methods (such as difference-in-difference (DiD) estimator or negative outcome control) have been proposed in this setting in the literature. However, these approaches require either restrictive assumptions on the data generating model or having access to at least two proxy variables.
We propose a method to estimate the causal effect using cross moments between the treatment, the outcome, and the proxy variable. In particular, we show that the causal effect can be identified with simple arithmetic operations on the cross moments if the latent confounder in linear SCM is non-Gaussian.
In this setting, DiD estimator provides an unbiased estimate only in the special case where the latent confounder has exactly the same direct causal effects on the outcomes in the pre-treatment and post-treatment phases. This translates to the common trend assumption in DiD, which we effectively relax.
Additionally, we provide an impossibility result that shows the causal effect cannot be identified if the observational distribution over the treatment, the outcome, and the proxy is jointly Gaussian.
 Our experiments on both synthetic and real-world datasets showcase the effectiveness
of the proposed approach in estimating the causal effect.
\end{abstract}

\section{Introduction}
Estimating the effect of a treatment (or an action) on an outcome is an important problem in many fields such as healthcare \cite{shalit2017estimating}, social sciences \cite{gangl2010causal}, and economics \cite{imbenscausal}. Randomized control trials are the gold standard to estimate causal effects. However, in many applications, performing randomized experiments are too costly or even infeasible, say due to ethical or legal concerns. Thus, estimating the causal effect from merely observational studies is one of the main topics of interest in causal inference. This problem has been studied extensively in two main frameworks, potential outcome (PO) framework \cite{rubin1974estimating} and structural causal model (SCM) framework \cite{pearl2009causality}. The main quantity of interest in PO framework is the individual-based response variable, i.e., the value of outcome for a specific individual in the population considering a particular value for the treatment. In SCM framework, a set of structural causal assignments are defined to describe the data generation mechanism among a set of variables. This set of assignments is often represented by a directed acyclic graph (DAG) to show the causal relationships among the variables in the model. It can be shown that the two frameworks are logically equivalent in the sense that any theorem in one can be translated to the other \cite{peters2017elements}.

Difference-in-Difference (DiD) \cite{lechner2011estimation} is one of the most frequently used non-experimental methods to estimate the effect of a treatment  by comparing the average of outcome before and after applying the treatment in a treatment and control group. In fact, 26 of 100 most cited papers published by the American Economic Review used some variant of DiD or two-way fixed effect (an extension to multi-group and multi-time slots) to estimate the causal effect \cite{de2022difference}. DiD is  an estimation process in PO framework for the setting where we have access to a population  partitioned into control and treatment groups. The goal is to estimate the effect of treatment $D$ on outcome $Y$ where $D$ is equal to one if a treatment is given to an individual and zero otherwise. It is also assumed that the value of the outcome is observed just before giving any treatment (this pre-treatment value is denoted by $Z$) and it can be seen  as a proxy variable for latent common causes of $D$ and $Y$. DiD method computes the causal effect by subtracting the difference of average outcome in two groups before applying treatment (i.e., $\E[Z|D=1]-\E[Z|D=0]$) from the one after the treatment (i.e., $\E[Y|D=1]-\E[Y|D=0]$). It can be shown the output of DiD is an unbiased estimate of the causal effect under some assumptions such as the parallel/common trend assumption which states that the outcome of the treatment group would have followed the same trend as the control group in the absence of the treatment (see \eqref{eq:common trend} for the exact definition).

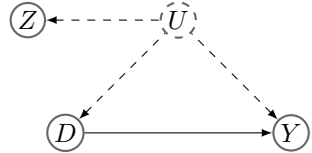
\begin{wrapfigure}{r} {0.5\textwidth}
    \begin{center}
        \begin{tikzpicture}[
            roundnode/.style={circle, draw=black!60,, fill=white, thick, inner sep=1pt},
            dashednode/.style = {circle, draw=black!60, dashed, fill=white, thick, inner sep=1pt},
            ]
            \node[roundnode]        (D)        at (-1.5, 0)                   {$D$};
            \node[roundnode]        (Y)        at (1.5, 0)                    {$Y$};
            \node[roundnode]        (Z)        at (-2, 1.5)                   {$Z$};
            \node[dashednode]       (U)        at (0, 1.5)                    {$U$};
            
            \draw[-latex] (D.east) -- (Y.west) node[midway,sloped,above] {} ;
            \draw[latex-, dashed] (D) -- (U) node[midway,sloped,above] {};
            \draw[latex-, dashed] (Y) -- (U) node[midway,sloped,above] {};
            \draw[latex-, dashed] (Z) -- (U) node[midway,sloped,above] {}; 
        \end{tikzpicture}
    \end{center}
    \caption{The suggested causal graph in SCM framework for the approaches in DiD and negative outcome control.}
    \label{fig: DiD graph}
\end{wrapfigure}

Although the initial setting of DiD is in PO framework, its counterpart in the SCM framework was considered in the negative outcome control approach \cite{sofer2016negative}. A negative outcome variable is a type of proxy variable that is not causally affected by the treatment. The causal graph in this approach is represented in Figure \ref{fig: DiD graph} where 
the unmeasured common cause of $D$ and $Y$ is represented by a latent variable $U$ and $D$ is not a cause of proxy variable $Z$.
The causal effect of $D$ on $Y$ cannot be identified from the observational distribution over $(D,Y,Z)$ because of the common confounder $U$. However, imposing further assumptions on the SCM, the causal effect of $D$  on $Y$ can become identified. Such assumptions include monotonicity \cite{sofer2016negative}, knowledge of the conditional probability $P(Z|U)$ \cite{kuroki2014measurement}, or having at least two proxy variables \cite{kuroki2014measurement,miao2018identifying,tchetgen2020introduction,cui2020semiparametric}, all of which may not hold in practice (see related work in Section \ref{sec:related work} for a more detailed discussion).
Recently,
\cite{salehkaleybar2020learning} considered linear SCMs with non-Gaussian exogenous noise \footnote{More precisely, at most one of the exogenous noises in the system can be Gaussian.} and proposed a method that can identify the causal effect for the causal graph in Figure \ref{fig: DiD graph} from the observational distribution over $(D,Y,Z)$. The proposed method is based on solving an over-complete independent component analysis (OICA) \cite{hyvarinen2001}. However given the landscape of the optimization problem, in practice, OICA can get stuck in bad local minima and return wrong results \cite{ding2019likelihood}. 

In this paper, we consider the setup in  causal graph in Figure \ref{fig: DiD graph} in linear SCM where we have access to a proxy variable $Z$.

We propose a ``Cross-Moment" algorithm that estimates the causal effect using cross moments between the treatment, the outcome, and the proxy variable. Our main contributions are as follows:

\begin{itemize}[leftmargin=*]
    \item We show that the causal effect can be identified correctly from the observational distribution if there exists $n\in \mathbb{N}$ such that for the latent confounder $U$, we have: $\E[U^n]\neq (n-1) \E[U^{n-2}]\E[U^2]$ (Theorem \ref{th:main}). Under additional mild assumption (Assumption \ref{assum: eq iff gaussian}), this condition implies that our proposed method can recover the causal effect when $U$ is non-Gaussian. Additionally, when the observational distribution is jointly Gaussian, we prove that it is impossible to identify the causal effect uniquely (Theorem \ref{th: impossibility result}).
    \item Unlike previous work \cite{salehkaleybar2020learning,adams2021identification} which requires solving an OICA problem, the proposed approach only performs simple arithmetic operations on cross moments. Therefore, it does not suffer from the drawbacks of  OICA such as getting stuck in bad local optima.
    \item We show that DiD estimator in general provides a biased estimate of the causal effect over the data generated by the linear SCM consistent with the causal graph in Figure \ref{fig: DiD graph} unless the latent confounder has exactly the same values of direct causal effects on the outcomes in the pre-treatment and post-treatment phases. Our proposed method  does not require such a strong restriction.   
\end{itemize}

The structure of the paper is as follows. In Section 2, we define the notation and provide some background on DiD estimator. In Section 3, we describe Cross-Moment algorithm and show that it recovers the true causal effect under mild assumptions on the distribution of the latent confounder. We also show that DiD estimator is in general biased if the data generative model follows a linear SCM. 
In Section 4, we review the related work. In Section 5, we evaluate the proposed algorithm experimentally and show its superior performance compared to the state of the art. Finally, we conclude the paper in Section 6.

\section{Preliminaries and Notations}
Throughout the paper, we denote random variables by capital letters and their realizations by small letters e.g., $X$ and $x$, respectively. Bold capital letters are used to specify a set of random variables and their realizations are denoted by small capital letters (e.g., $\X$ and $\x$). 

 A SCM $\M$ over a set of random variable $\V$ is defined by a set of assignments $\{X := f_X^{\M}(\Pa{X}{\G}, \epsilon_X)\}_{X\in \V}$, where $\epsilon_X$ is the exogenous noise corresponding to $X$ and $\Pa{X}{\G} \subseteq \V$. It is assumed that the exogenous noises are mutually independent.  Let us denote by $\mathbf{O}$ and $\U$, the set of observed and unobserved variables in $\V$, respectively. Note that $\V = \mathbf{O}\cup \U$ and $\O \cap \U=\emptyset$. 
 
 The set of assignments in SCM $\M$ is  commonly represented by a DAG. Let $\G=(\V, \mathbf{E})$ be a DAG with the set of vertices $\V$ and set of edges $\mathbf{E}$. For ease of notation, we use the notation of $\V$ for the set of vertices in the graph. We also use the term ``vertex" and ``random variable" interchangeably. 
Each vertex in the graph represents some random variable and each direct edge shows a direct causal relationship between a pair of random variables. 
In particular, we say that $X$ is a parent of $Y$ or, equivalently, $Y$ is a child of $X$ if $(X, Y) \in \mathbf{E}$.  We define  $\Pa{X}{\G}$ as a set of all parents  of $X$ in graph $\G$.

\subsection{Difference-in-Difference (DiD)}
\label{sec: Preliminaries-DiD}
Difference-in-difference (DiD) was proposed in the PO framework in order to estimate the causal effect from observational studies under some assumptions. In this framework, the population under study is divided into control and treatment groups and only individuals in the treatment group receive the treatment. In particular, the treatment variable $D$ represents treatment assignment which is equal to $1$ if the treatment was given and $0$ otherwise. Let $Y(0)$ and $Y(1)$ be two random variables representing
the outcome under treatment value $D=0$ and $D=1$, respectively. Denote the value of the outcome right before administering the treatment by $Z$ and assume it is measurable.
Our goal is to obtain the average causal effect in the treatment group $\mathbb{E}[Y(1)-Y(0)|D=1]$.
DiD estimate of the average causal effect equals:
\begin{equation}
    (\E[Y|D=1]-\E[Y|D=0]) - (\E[Z|D=1]-\E[Z|D=0]).
    \label{eq:DiD}
\end{equation}
This quantity is an unbiased estimate of the average causal effect as long as the following assumptions hold.
\begin{itemize}[leftmargin=*]
    \item Stable Unit Treatment Value Assumption (SUTVA): 
    \begin{equation*}
    \begin{split}
        Y = DY(1) + (1-D)Y(0).
    \end{split}
    \end{equation*}

    \item Common trend assumption:
    \begin{equation}
    \label{eq:common trend}
    \E[Y(0)-Z(0)|D=1]=\E[Y(0)-Z(0)|D=0].
    \end{equation}
\end{itemize}

SUTVA states that the potential outcome for each individual is not related to the treatment value of the other individuals. The common trend assumption states that there would be the same ``trend'' in both groups in the absence of treatment which allows us to subtract group-specific means of the outcome in estimating the average causal effect in \eqref{eq:DiD}.

\section{Methodology: Cross-Moment Algorithm }

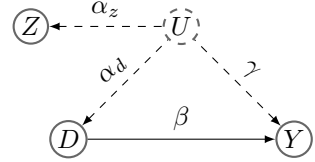
\begin{wrapfigure}{r}{0.5\textwidth}
  \begin{center}
    \begin{tikzpicture}[
            roundnode/.style={circle, draw=black!60,, fill=white, thick, inner sep=1pt},
            dashednode/.style = {circle, draw=black!60, dashed, fill=white, thick, inner sep=1pt},
            ]
            \node[roundnode]        (D)        at (-1.5, 0)                   {$D$};
            \node[roundnode]        (Y)        at (1.5, 0)                    {$Y$};
            \node[roundnode]        (Z)        at (-2, 1.5)          
              {$Z$};
            \node[dashednode]       (U)        at (0, 1.5)                    {$U$};
            
            \draw[-latex] (D.east) -- (Y.west) node[midway,sloped,above] {$\beta$} ;
            \draw[latex-, dashed] (D) -- (U) node[midway,sloped,above] {$\alpha_d$};
            \draw[latex-, dashed] (Y) -- (U) node[midway,sloped,above] {$\gamma$};
            \draw[latex-, dashed] (Z) -- (U) node[midway,sloped,above] {$\alpha_z$};
        \end{tikzpicture}
  \end{center}
  \caption{The considered causal graph with linear assignments in the SCM framework.}
  \label{fig: main graph}
\end{wrapfigure}

In this section, we propose Cross-Moment algorithm 
to estimate the causal effect of treatment $D$ on outcome $Y$.
Throughout this section, we consider linear SCMs, i.e., each random variable in SCM $\M$ is a linear combination of its parents and its corresponding exogenous noise.  More precisely, the linear assignments in $\mathcal{M}$ for the causal graph in Figure \ref{fig: main graph} are:
\begin{equation}
\label{eq: main SEM}
\begin{split}
    U &:= \epsilon_u,\\
    Z &:= \alpha_z U + \epsilon_z =\alpha_z \epsilon_u + \epsilon_z, \\
    D &:= \alpha_d U + \epsilon_d = \alpha_d \epsilon_u + \epsilon_d, \\
    Y &:= \beta D + \gamma U +\epsilon_y =(\alpha_d \beta + \gamma) \epsilon_u + \beta \epsilon_d + \epsilon_y,
\end{split}
\end{equation}
Without loss of generality, we assume that $\epsilon_u$, $\epsilon_y$, $\epsilon_z$, $\epsilon_d$ are arbitrary random variables with zero mean. 

Moreover, we assume that the only observed random variables are given by  $\O=\{D,Y,Z\}$.  Our goal is to identify $\beta$ (the causal effect of $D$ on $Y$) from the distribution over the observed variables $\O$. We consider SCMs that satisfy the following assumption.
\begin{assumption}\label{assum: Var(eps_d)>0}
    In the linear SCM given by \eqref{eq: main SEM},  $\alpha_z \neq 0$ and $\Var{\epsilon_d}>0$.
\end{assumption}
Assumption \ref{assum: Var(eps_d)>0} is necessary for identifying the causal effect. In particular, if $\alpha_z= 0$, the directed edge from $U$ to $Z$ is removed and the causal effect cannot be identified even if all the exogenous noises are non-Gaussian as shown in \cite{salehkaleybar2020learning}. Moreover, if $\Var{\epsilon_d}=0$, then $\epsilon_d$ is zero almost surely (as we assumed that all the exogenous noises are mean zero). In this case, we can construct another SCM $\mathcal{M}'$ which encodes the same observational distribution as our original SCM but results in a different value of the causal effect of $D$ on $Y$ compared to the original SCM. More specifically, in this SCM, we delete
the directed edge from $U$ to $Y$ and change the structural assignment of $Y$ to $Y:=(\beta+\gamma/\alpha_d)D+\epsilon_y$.  Hence, the assumption $\Var{\epsilon_d}>0$ is necessary for the unique identification of the causal effect.

Under Assumption \ref{assum: Var(eps_d)>0}, it can be shown that:
\begin{equation}
    \beta = \frac{
    \Cov{D}{Y} - \frac{\alpha_d}{\alpha_z}\Cov{Y}{Z}
    }{\Var{D} - \frac{\alpha_d}{\alpha_z}\Cov{D}{Z}},
    \label{eq:compute beta}
\end{equation}
where $\Cov{A}{B}$ denotes the covariance of random variables $A$ and $B$ and $\Var{A}$ is the variance of $A$. $\beta$ is identifiable as long as the ratio $\alpha_d/\alpha_z$ is known as we can obtain $\Cov{D}{Y},\Cov{Y}{Z},\Var{D},$ and $\Cov{D}{Z}$ from the observational distribution. In the sequel, we will show how this ratio can be learnt as long as $\epsilon_u$ has bounded moments.

\begin{assumption}\label{assum: finite moments}
   For all $n\in \mathbb{N}$, assume that: $\E[\epsilon_u^n] < \infty$.
\end{assumption}
When the bounded moment assumption of \ref{assum: finite moments} holds, the following theorem provides an approach for recovering  $\alpha_d/\alpha_z$.

\begin{theorem}
\label{th:main}
    For variables $Z$ and $D$ as defined in \eqref{eq: main SEM}, under Assumptions \ref{assum: finite moments},
    $\frac{\alpha_d}{\alpha_z}$ can be determined uniquely if $\exists n\in \mathbb{N}$ such that:
        \begin{equation}
        \label{eq:cond in Theorem}\E\left[\hat{\epsilon}_u^{n}\right] \neq (n-1)\E\left[\hat{\epsilon}_{u}^{n-2}\right]\E\left[\hat{\epsilon}_{u}^2\right],
        \end{equation}
        where $\hat{\epsilon}_u=\sqrt{\alpha_d\alpha_z}\epsilon_{u}$.

\end{theorem}

The detailed proof of the Theorem \ref{th:main} is provided in the Appendix \ref{app: technical proofs}.

It is interesting to see for what families of distributions, the condition in Theorem \ref{th:main} is satisfied. Assume \eqref{eq:cond in Theorem} is not satisfied. Recall that from the definition of SCM \eqref{eq: main SEM},  $\E[\hat{\epsilon}_u]=0$ and $\E[\left(\hat{\epsilon_u}\right)^2]=\E[DZ]$. These in a combination with $\E\left[\hat{\epsilon}_u^{n}\right] = (n-1)\E\left[\hat{\epsilon}_{u}^{n-2}\right]\E\left[\hat{\epsilon}_{u}^2\right]$ for any $n\in \mathbb{N}$ determine uniquely all the moments of $\hat{\epsilon}_u$. More specifically, recursively solving for $\E\left[\hat{\epsilon}_u^{n}\right]$  we have  $\E\left[\hat{\epsilon}_u^{n}\right]=(n-1)!!\E[\left(\hat{\epsilon_u}\right)^2]$ for even $n\geq 1$ and $\E\left[\hat{\epsilon}_u^{n}\right]=0$ for odd $n\geq 1$ where $n!!$ denotes double factorial. Double factorial notation $n!!$ denotes the product of all numbers from 1 to $n$ with  the same parity as $n$. Specifically the moments of Gaussian distribution satisfy the aforementioned moment equation. Therefore   when $\epsilon_u$ is Gaussian, we cannot identify the causal effect. Under some mild technical assumption on $\epsilon_u$ (see Assumption \ref{assum: eq iff gaussian} in the following), we can prove that the moments of $\epsilon_u$ \textit{uniquely} determine its distribution. As a result, as long as $\epsilon_u$ is non-Gaussian, we can identify the causal effect. 



\begin{assumption}\label{assum: eq iff gaussian}
    We assume that there exists some $s>0$ such that the power series $\sum_k \E[\epsilon_u^k] r^k/k!$ converges for any $0<r<s$. 
\end{assumption}


\begin{corollary}
Under Assumptions \ref{assum: Var(eps_d)>0}, \ref{assum: finite moments} and \ref{assum: eq iff gaussian}, the causal effect $\beta$ can be recovered uniquely as long as $\epsilon_u$ is not Gaussian. 
\label{cor:main}
\end{corollary}

In \cite{salehkaleybar2020learning}, it was shown that $\beta$ can be recovered as long as \textit{all} exogenous noises are non-Gaussian. Therefore, our result relaxes the restrictions on the model in \cite{salehkaleybar2020learning} by allowing $\epsilon_D,\epsilon_Y,\epsilon_Z$ to be Gaussian.

Based on Theorem \ref{th:main}, we present Cross-Moment algorithm in Algorithm \ref{algo: compute beta} that computes coefficient $\beta$ from the distribution over the observed variables $Z, D, Y$. Algorithm \ref{algo: compute beta} is comprised of two functions $\textbf{\textit{GetRatio}}$ and $\textbf{\textit{GetBeta}}$. In the proof of Theorem \ref{th:main}, we show that $ |\alpha_d/\alpha_z| = (\text{num}/\text{den})^{1/(n-2)}$ for the smallest $n$ such that $\text{den}\neq 0$ where $\text{num}$ and $\text{den}$ are defined in lines 6 and 7 of function $\textbf{\textit{GetRatio}}$, respectively. Moreover, $\E[DZ]$ has the same sign as $\alpha_d/\alpha_z$ and we can recover the sign of the ratio $\alpha_d/\alpha_z$ from $\E[DZ]$. Thus, in lines 8-10, for the smallest $n$ such that $\text{den}\neq 0$, we obtain the ratio $\alpha_d/\alpha_z$ and then use it in function $\textbf{\textit{GetBeta}}$ to recover $\beta$.

\begin{algorithm}[t]
    \caption{Cross-Moment algorithm}
    \label{algo: compute beta}
    \begin{algorithmic}[1]
        \STATE $\textbf{Function \textit{GetBeta}} (D, Z, Y)$
        \STATE $\text{ratio} := \textbf{\textit{GetRatio}}(D, Z)$
        \STATE $\beta := (\E[DY] - \text{ratio}\cdot\E[YZ])/(\E[D^2] - \text{ratio}\cdot\E[DZ])$
        \STATE $\textbf{Return } \beta$
    \end{algorithmic}
    \hrulefill
    \begin{algorithmic}[1]
        \STATE $\textbf{Function \textit{GetRatio}}(D, Z)$
        \STATE $\text{findRatio} := \mathbf{False}$
        \STATE $n := 2$
        \WHILE{$\text{findRatio}\neq \mathbf{True}$}
            \STATE $n := n+1$
            \STATE $\text{num} :=  \E[D^{n-1}Z] - (n-1)\E[D^{n-2}] \E[DZ]$
            \STATE $\text{den} := \E[Z^{n-1}D] - (n-1)\E[Z^{n-2}] \E[DZ]$
            \IF{$\text{den} \neq 0$}
                \STATE $\text{ratio} := \textbf{\text{sign}}( \E[DZ])\left|(\frac{\text{num}}{\text{den}})^{1/(n-2)}\right|$
                \STATE $\text{findRatio} := \mathbf{True}$
            \ENDIF
        \ENDWHILE
        \STATE $\textbf{Return: } \text{ratio}$
    \end{algorithmic}
\end{algorithm}

\subsection{Impossibility Result}
In the previous section, we showed that the causal effect $\beta$ can be identified if the distribution of latent confounder is non-Gaussian. Herein, we show that no algorithm can learn $\beta$ uniquely if the observed variables are jointly Gaussian in any linear SCM defined by \eqref{eq: main SEM} satisfying the following assumption. 

\begin{assumption}\label{assum: Var(eps_z)>0}
    In the linear SCM defined by \eqref{eq: main SEM}, 
    $\alpha_d \neq 0$, $\gamma \neq 0$ and $\Var{\epsilon_z}>0$.
\end{assumption}
\begin{theorem}\label{th: impossibility result}
    Suppose that the observed variables in linear SCM defined by \eqref{eq: main SEM} are jointly Gaussian.
    Under Assumptions \ref{assum: Var(eps_d)>0}, \ref{assum: finite moments} and \ref{assum: Var(eps_z)>0}, the total causal effect $\beta$ cannot be identified uniquely.
\end{theorem}
The proof of the Theorem \ref{th: impossibility result} appears in the Appendix \ref{app: technical proofs}. The key idea in the proof is to show that there exist two linear SCMs that encode the same observational distribution and are consistent with the causal graph in Figure \ref{fig: main graph} but the causal effect of $D$ on $Y$ has two different values in these two models.

Note that it is  known that the causal structure is not identifiable in a linear SCM with Gaussian exogenous noises \cite{peters2017elements}. Our impossibility result here is different from the non-identifiability result in linear Gaussian models. Specifically, in the linear Gaussian models, the goal is to recover all the coefficients in the linear SCM from the observational distribution. In our setting, we have additional knowledge of the exact DAG (restriction on the form of the linear SCM in \eqref{eq: main SEM}), and the goal is to identify a specific coefficient (i.e., $\beta$) from the linear SCM. Therefore, we have more constraints on the model and need to infer less information about it. Still,  we show that the target coefficient $\beta$ cannot be determined in the causal graph in Figure \ref{fig: main graph}  if the observed variables are jointly Gaussian.

\subsection{Bias in DiD Estimator}
Suppose that the data is generated from a linear SCM consistent with the causal graph in Figure \ref{fig: main graph}. We show that DiD estimator is biased except when the latent variable $U$ has the exact same direct causal effect on $Z$ that it has on $Y$, i.e., $\alpha_Z=\gamma$. Our Cross-Moment algorithm identifies the true causal effect without any such restrictive assumption on the coefficients of the linear SCM. 

DiD estimator is given by the following linear regression \cite{lechner2011estimation}:
\begin{equation}\label{eq: TWFE regression}
    \hat{Y}= \hat{\beta}_1 T + \hat{\beta}_2D +\hat{\beta} DT,
\end{equation}
where $\hat{\beta}_1$, $\hat{\beta}_2$, and $\hat{\beta}$ are the regression coefficients and $T$ is a binary variable that equals zero for the pre-treatment phase and equals one otherwise. In the pre-treatment phase, $Z$ (i.e., the outcome before the treatment) is predicted as $\hat{\beta}_2D$ and in the post-treatment phase, $Y$ (the outcome after giving treatment to the treatment group) is predicted accordingly as $\hat{\beta}_1+(\hat{\beta}+\hat{\beta}_2)D$. In order to obtain the regression coefficients, the expectation of squared residuals over the population is minimized as follows (see Appendix \ref{app: DiD} for the derivations of the following minimization problem and subsequent equations in this section):
\begin{equation*}
\min_{\hat{\beta}_1, \hat{\beta}_2, \hat{\beta}}\E[(Z-\hat{\beta}_2D)^2]+\E[(Y-\hat{\beta}_1-(\hat{\beta}+\hat{\beta}_2)D)^2].
\end{equation*}
This results in the following regression coefficients:
\begin{equation*}
    \hat{\beta}_1=0, \quad \hat{\beta}_2= \frac{\E[ZD]}{\E[D^2]},\quad\hat{\beta}=\frac{\E[YD]-\E[ZD]}{\E[D^2]}.
\end{equation*}
DiD estimator returns $\hat{\beta}$ in the above equation as the estimation of causal effect which is equal to:
\begin{equation}
    \hat{\beta}= \beta +\frac{\alpha_d(\gamma-\alpha_z)\E[U^2]}{\E[D^2]}.
    \label{eq:DiD linear SCM}
\end{equation}
Thus, $\hat{\beta}$ is an unbiased estimate of $\beta$ only when $\gamma=\alpha_z$. In other words, latent variable $U$ should have the same direct causal effect on $Z$ and $Y$. This is akin to the so-called common trend assumption which says that the average
natural drift (here, the effect of $U$) is assumed to be the same across both the control and treatment groups. In summary, whenever the common trend assumption is violated, the DiD estimator is biased.

\section{Related work}
\label{sec:related work}
In the past few years, there has been a growing interest in the literature to exploit proxy variables to de-bias the effect of latent confounders. A special type of such proxy variable is the so-called negative outcome which is a variable known not to be causally affected by the treatment \cite{lipsitch2010negative}.
For instance, the variable $Z$ in Figure \ref{fig: DiD graph} may be considered as negative outcome. In fact, \cite{sofer2016negative} interpreted DiD  as a negative outcome control approach and proposed a method inspired by change-in-change \cite{athey2006identification} to identify the causal effect under the assumption that $Y(0)$ and $Z$ are monotonic increasing functions of latent confounders and some observed covariates.

\cite{kuroki2014measurement} considered three settings in causal inference with proxy variables: 1- There exists only one proxy variable such as $Z$ as a negative outcome. In this case, for discrete finite variables $Z$ and $U$, they showed that the causal effect can be identified if $\Pr(Z|U)$ is known from some external studies such as pilot studies. 2- Two proxy variables, for instance $Z$ and $W$ are considered where $U$, $Z$, and $W$ are all discrete finite variables and $Z$ does not have a directed path to $D$ or $Y$. It has been shown that the causal effect is identifiable under some assumptions on the conditional probabilities of $\Pr(Y|D,U)$ and $\Pr(Z,W|X)$. In the setting, it is not necessary to know $\Pr(Z|U)$ but two proxy variables are required to identify the causal effect. 3- In linear SCM, \cite{kuroki2014measurement} showed that $\beta$ (the average causal effect of $D$ on $Y$) can be recovered using two proxy variables. 
Later, \cite{miao2018identifying} also considered a setting with two proxy variables $Z$ and $W$. Unlike the second setting in \cite{kuroki2014measurement}, here, $Z$ and $W$ can be parents of $D$ and $Y$, respectively. For the discrete finite variables, they showed that the causal effect can be identified if the matrix $P(W|Z,D=d)$ is invertible. Moreover, they provided the counterpart of this condition for continuous variables. \cite{shi2020multiply} extended the identification result in \cite{miao2018identifying}, with a weaker set of assumptions. Still, they required two proxy variables to identify the causal effect. Based on the results in \cite{miao2018identifying}, \cite{tchetgen2020introduction} introduced a proximal causal inference in PO framework. More recently, \cite{cui2020semiparametric} provided an alternative proximal identification result to that of \cite{miao2018identifying}, again when two proxy variables were avaialble. 

In linear SCMs, to the best of our
knowledge, the methods that can identify the causal effect with only one proxy variable in Figure \ref{fig: main graph} are based on solving an OICA problem. In particular,
\cite{salehkaleybar2020learning} considered linear SCM with non-Gaussian exogenous noises in the presence of latent variables. They showed that under some structural conditions, the causal effects among observed variables can be identified and the causal graph in Figure \ref{fig: main graph} satisfies such structural conditions. However, the proposed method requires solving an OICA and the existing methods for solving such a problem might get stuck in bad local optima. Very recently, \cite{adams2021identification} provided two graphical conditions for the same setting in \cite{salehkaleybar2020learning} which are necessary for the identification of the causal structure. These conditions are closely related to the sparsity of the causal graphs. For the causal graph in Figure \ref{fig: main graph}, the method proposed in \cite{adams2021identification} for estimating the causal effect is the same as the one in \cite{salehkaleybar2020learning} and thus has the same drawback.



In PO framework, the setting of having just a pre-treatment phase and a post-treatment phase can be generalized to the case with multiple time slots in the panel data model \cite{athey2021matrix}. In this paper, we mainly focus on the setting with two groups and two time slots but one can also study the extensions of the current work for other settings in the panel data model described in the following. Consider two $N\times T$ matrices $\Y$ and $\D$ where $N$ is the number of individuals in the population and $T$ is the number of time slots. Assume that only the outcome for some individuals and time slots is observable. In particular: $Y_{it}=(1-D_{it})Y_{it}(0)+D_{it}Y_{it}(1)$, where the realized outcome for individual $i$ at time slot $t$ is denoted by $Y_{it}(D_{it})$. DiD method has been proposed for the case $T=2$, i.e., two time slots (pre-treatment and post-treatment phases). 
In the literature, other cases have been also studied for various assumptions on matrix $\Y$. For instance, in unconfounded case \cite{rosenbaum1983central,imbenscausal}, the number of individuals is much larger than the number of time slots ($N\gg T$), and the treatment is provided only at the last time slot.
Another setting is that of synthetic control \cite{abadie2003economic,abadie2010synthetic,abadie2015comparative,doudchenko2016balancing} where $T\gg N$. In this setting, there is a single treated individual (suppose individual $N$) and the goal is to estimate its missing potential outcomes for any $t\in [T_0,T]$ after administering the treatment at time $T_0$.
The last setting considers $N\approx T$ and a two-way-fixed-effect (TWFE) regression model has been proposed to estimate the causal effect (see for a survey on TWFE in \cite{de2020two}).
It is noteworthy that TWFE estimator is equivalent to DiD estimator for two groups and two time slots.

\section{Experiments}
In this section, we first evaluate our algorithm on synthetic data and compare it to DiD estimator and as well as the related work in \cite{kuroki2014measurement} which estimates the causal effect in linear SCMs with two proxy variables. Further, we apply our algorithm to a real dataset provided by \cite{card1993minimum}.

 \subsection{Synthetic data}
 \label{sef: Synthetic data}

 We generated samples according to the SCM in \eqref{eq: main SEM} and with all the exogenous noises distributed according to exponential distribution. Note that the distribution of $\epsilon_u$, i.e., the exponential distribution satisfies Assumptions  \ref{assum: finite moments} and \ref{assum: eq iff gaussian}. Therefore $\beta$ is identifiable according to the Corollary \ref{cor:main}.

Given the observational data, we estimated the value of $\beta$ from the following four approaches:
\begin{enumerate}
    \item Cross-Moment algorithm (proposed in this work).
    \item DiD estimator of \eqref{eq: TWFE regression}.
    \item A simple linear regression model based on the following equation:
        $\hat{Y} = \alpha Z + \hat{\beta} D.$
    \item Causal effect estimate for linear SCM with two proxy variables (proposed in  \cite{kuroki2014measurement}). In the experiments, we call this estimate ``two-proxy'' method.
\end{enumerate}
It is noteworthy that we also evaluated the method in \cite{salehkaleybar2020learning} which uses OICA as a subroutine. Unfortunately, the performance was too poor to be included.

\begin{figure}
    \begin{subfigure}[t]{0.45\textwidth}
        \includegraphics[width=7cm]{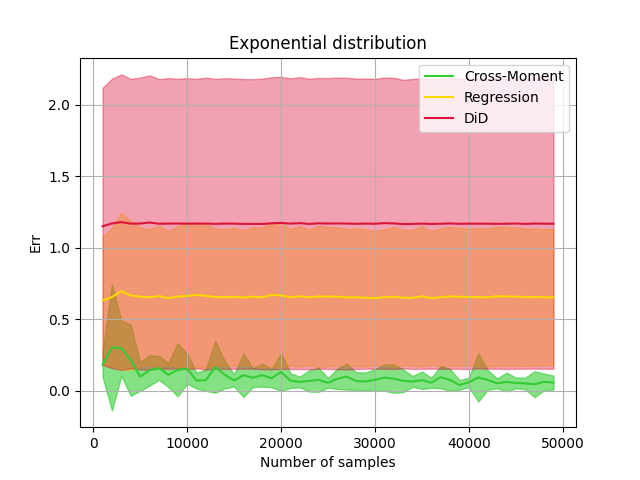}
        \caption{The average relative error of Cross-Moment, DiD estimator, and simple linear regression with one proxy variable.}
        \label{fig: results exponential samples}
    \end{subfigure}
    \hfill
    \begin{subfigure}[t]{0.45\textwidth}
        \includegraphics[width=7cm]{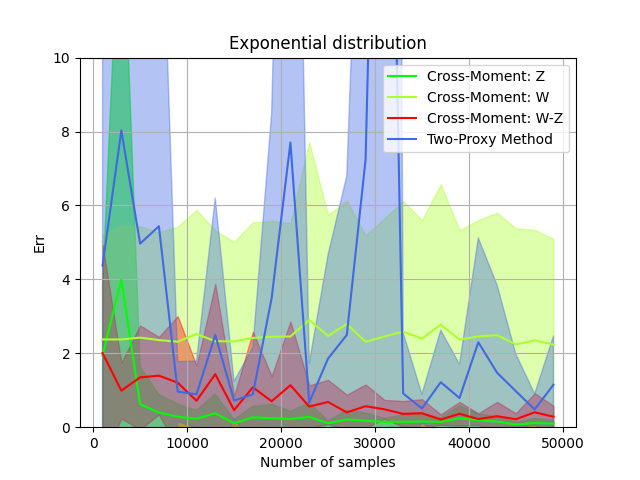}
        \caption{The average relative error of three variants of Cross-Moment algorithm and the two-proxy method in \cite{kuroki2014measurement} when we have access to two proxy variables.}
        \label{fig: results 2 proxies}
    \end{subfigure}
    \caption{The performance measure $err$ against the number of samples. Colored regions show the standard deviation of the $err$.}
\end{figure}

For each sample size, we sampled parameters $\alpha_z$, $\alpha_d$, $\beta$, $\gamma$ randomly  and then generated the samples of $Z$, $D$, $Y$ accordingly to \eqref{eq: main SEM} (More details regarding the data generation mechanism can be found in Appendix \ref{app: experiments}). We ran an experiment $10$ times and reported the 
the average relative error for each value of sample size:
    $err = \E\left[\left|\frac{\beta-\hat{\beta}}{\beta}\right|\right].$

Figure \ref{fig: results exponential samples}, depicts the performances of Cross-Moment algorithm, DiD estimator, and the simple linear regression when we have access to only one proxy variable. The colored region around each curve shows the empirical standard deviation of $|(\beta-\hat{\beta})/\beta|$.
Cross-Moment algorithm outperforms the other two methods significantly. In fact, DiD estimator is biased if $\alpha_z\neq\gamma$ which occurs with measure one as $\alpha_z$ and $\gamma$ are generated randomly. Moreover, DiD estimate is no better than simple linear regression if $\gamma$ is not close to $\alpha_z$. In the literature, it has been noted that the parallel trend assumption (in linear SCM, this assumption is equivalent to the condition $\alpha_z=\gamma$) is violated if the scale of the proxy variable $Z$ and outcome variable $Y$ are different which can be the case in many practical applications \cite{lechner2011estimation}. 

We compared Cross-Moment with the two-proxy method in \cite{kuroki2014measurement} when we have access to two proxy variables. In particular, we assumed that there is an additional proxy variable $W$ such that $W := \alpha_w U + \epsilon_w$. For Cross-Moment algorithm, we considered three versions: \MakeUppercase{\romannumeral 1} - ``Cross-Moment: Z'', which estimates the causal effect by  using only proxy variable $Z$ (we denote this estimate by $\beta_Z$), \MakeUppercase{\romannumeral 2} - ``Cross-Moment: W'', which estimates $\beta$ from only proxy variable $W$ (which we denote the estimate by $\beta_W$), \MakeUppercase{\romannumeral 3} - ``Cross-Moment: W-Z'', which estimates $\beta$ from aggregating the estimates of the methods \MakeUppercase{\romannumeral 1} and \MakeUppercase{\romannumeral 2}. In particular, ``Cross-Moment: W-Z'' uses bootstrapping method (Monte Carlo algorithm for case resampling \cite{efron1994introduction}) to estimate  the variances of estimates $\beta_Z$ and $\beta_W$, denoted by $\sigma^2_{\beta_Z}$ and $\sigma^2_{\beta_W}$, respectively. Subsequently, $\beta$ is estimated by combining two estimates $\beta_Z$ and $\beta_W$ with an inverse-variance weighting scheme \citep{sinha2011statistical} where we give a higher weight to the estimate with the lower variance: $\frac{\sigma^2_{\beta_Z}}{\sigma^2_{\beta_Z}+\sigma^2_{\beta_W}}\beta_W+\frac{\sigma^2_{\beta_W}}{\sigma^2_{\beta_Z}+\sigma^2_{\beta_W}}\beta_Z$. When
$\Var{\epsilon_w}/\Var{W}$ and $\Var{\epsilon_z}/\Var{Z}$ are small, the causal effect can be estimated with a low estimation error from either $Z$ or $W$ as they contain low noise versions of the latent confounder $U$. In our experiments, we considered the case where one of the proxy variables (herein, $W$) is too noisy but not the other one. Specifically, we choose$\Var{\epsilon_w}/\Var{\epsilon_u}=10$  and $\Var{\epsilon_z}/\Var{\epsilon_u}=0.1$. Figure \ref{fig: results 2 proxies} illustrates the performances of the three aforementioned variants of Cross-Moment algorithm  and the two-proxy method in \citep{kuroki2014measurement}. ``Cross-Moment: Z'' has the best performance since it uses $Z$ with less noise as the proxy variable. Moreover, ``Cross-Moment: W-Z'' has a comparable performance by combining the estimates of  $\beta_Z$ and $\beta_W$. The two-proxy estimate does not exhibit robustness and has a large average relative error for various values of sample size.

\subsection{Minimum Wage and Employment Dataset}
 We evaluate our method on the real data  which contains information about fast-food stores (Burger King, Roy Rogers, and Wendy's stores) in New Jersey and Pennsylvania in 1992, and some details about them such as minimum wage, product prices, open hours, etc  \cite{card1993minimum}. The goal of study was to estimate the effect of the  in minimum wage in New Jersey from \$ 4.25 to \$ 5.05 per hour on the employment rate. 

 The data was collected by interviews in two waves, before and after the rise in the minimum wage.
 The information was gathered from $410$ restaurants with similar average food prices, store hours, and employment levels. In this experiment, stores from Pennsylvania are treated as a control group and stores from New Jersey are considered as the treatment group. 
 We defined employment level $Y$ as 
     $Y = Y_{p} + \frac{1}{2}Y_{f},$
 where $Y_{p}$ is a number of employees working part-time and $Y_{f}$ is a number of employees working full-time.  

 First, we reproduced the results in \cite{card1993minimum}.  We considered an extended version of TWFE model \cite{card1993minimum}:
 \begin{equation*}
     \hat{Y} = \hat{\beta}_1 T + \X^T\hat{\alpha} + \hat{\beta}_2D +\hat{\beta} DT,
 \end{equation*}
where $\hat{Y}$ is the estimate of number of employees in the store, $T$ is a binary variable that equals $0$ prior to raising the minimum wage and equals to $1$ after the raise. $D$ is equal to $0$ if the store is in Pennsylvania and equal to $1$ if the store is in New Jersey.  $\X$ is a vector that contains additional information such as the opening hours, product prices, etc. $\hat{\alpha}$ is also a vector of parameters corresponding to the vector $\X$.
We dropped all the stores from the dataset that contain $\text{NaN}$ values after which $246$ restaurants were left.
\begin{table}
\centering
\begin{tabular}{ |p{2.3cm}||p{2.3cm}|p{2.3cm}|}
 \hline
 &   TWFE   &   Cross-Moment   \\
 \hline
 With $\X$           &   2.68    &    2.68   \\
 \hline
 Without $\X$        &   3.24    &    4.03   \\
 \hline
\end{tabular}
 \caption{Causal effect estimation of minimum wage on employment level in the real dataset in \cite{card1993minimum}.}
 \label{table: results real-data}
\vspace{-0.4cm}
\end{table}
The estimate of $\beta$ computed by TWFE is given in the first row of Table \ref{table: results real-data}. According to \cite{card1993minimum}, the estimate of $\beta$ is equal to $2.76$. The difference in estimation is due to the slight difference in the features of the vector $\X$, i.e., \cite{card1993minimum} added a few additional manually computed features to $\X$.

For the Cross-Moment algorithm, in order to incorporate the features $\X$ in estimating $\beta$, we first regressed $Y$ on $\X$ and then used $ Y - \X\hat{\alpha}$ instead of $Y$ as the outcome.
The result of applying Cross-Moment algorithm to this newly defined outcome is given in the first row of Table \ref{table: results real-data} and  is very close to the estimate by TWFE.

Finally, we assumed that the additional information $\X$ gathered during the interview is not available. Then TWFE model for the employment level takes the following form
\begin{equation*}
         \hat{Y}= \hat{\beta}_1 T + \hat{\beta}_2D +\hat{\beta} DT.
\end{equation*}
We used the previously pre-processed dataset from but dropped the columns corresponding to $\mathbf{X}$. Subsequently, we applied TWFE and Cross-Moment method to estimate $\beta$. The respective estimates appear in the second row of Table \ref{table: results real-data}, which stipulates the rise in the minimum wage had a positive effect on the employment level.

\section{Conclusion}
\vspace{-0.2cm}
We considered the problem of estimating the causal effect of a treatment on an outcome a linear SCM where we have access to a proxy variable for the latent confounder of the treatment and the outcome. This problem has been studied in both PO framework (such as DiD estimator) and SCM framework (such as the negative outcome control approach). We proposed a method that uses cross moments between the treatment, the outcome, and the proxy variable and recovers the true causal effect if the latent confounder is non-Gaussian. We also showed that the causal effect cannot be identified if the joint distribution over the observed variable are Gaussian.  Unlike previous work which requires solving an OICA problem, Our performs simple arithmetic operations on the cross moments. We evaluated our proposed method on synthetic and real datasets. Our experimental results show  the proposed algorithm has remarkable performance for synthetic data and  provides consistent results with previous studies on the real dataset we tested on.

\bibliography{main}

\newpage
\appendix
\onecolumn

\section{Technical Proofs}
\label{app: technical proofs}
\begin{customthm}{\ref{th:main}}
    For the variables $Z$ and $D$ as defined in \eqref{eq: main SEM}, under Assumptions \ref{assum: finite moments},
    $\frac{\alpha_d}{\alpha_z}$ can be determined uniquely if $\exists n\in \mathbb{N}$ such that:
    \begin{equation}
    \E\left[\hat{\epsilon}_u^{n}\right] \neq (n-1)\E\left[\hat{\epsilon}_{u}^{n-2}\right]\E\left[\hat{\epsilon}_{u}^2\right],
    \end{equation}
    where $\hat{\epsilon}_u=\sqrt{\alpha_d\alpha_z}\epsilon_{u}$.
\end{customthm}
\begin{proof}
    Let $\alpha := \sqrt{\frac{\alpha_d}{\alpha_z}}$. Please note that $\alpha$ can be a complex number. Then, $D$ and $Z$ in \eqref{eq: main SEM}  can be rewritten as:
    \begin{equation*}
        \begin{split}
            & D = \alpha \hat{\epsilon}_u + \epsilon_d,\\
            & Z = \frac{1}{\alpha} \hat{\epsilon}_u + \epsilon_z.
        \end{split}
    \end{equation*}
    
     We prove by induction that:
     \begin{equation}
      \label{eq: prop norm dist}\E\left[\hat{\epsilon}_u^{n}\right] = (n-1)\E\left[\hat{\epsilon}_{u}^{n-2}\right]\E\left[\hat{\epsilon}_{u}^2\right],   
     \end{equation}
     holds for any $n \in \mathbb{N}$ or $\frac{\alpha_d}{\alpha_z}$ ($\alpha^2$ equivalently) can be identified uniquely from $D$ and $Z$. 
    
    \textit{Base of induction.}
    It is easy to verify that for $n=2$, \eqref{eq: prop norm dist} holds.
    
    \textit{Induction step.} Assume that $n\geq 2$  and  \eqref{eq: prop norm dist} holds for all $k<n$. Then we prove either it also holds for $n$ or $\alpha_d/\alpha_z$ can be identified uniquely.
    We have
    \begin{align} \label{eq: D^(n-1)Z}
        & \E[D^{n-1}Z] = \E[(\alpha\hat{\epsilon}_{u}+\epsilon_{d})^{n-1})\frac{1}{\alpha}\hat{\epsilon}_{u}] = \\
        \notag
        & \E[\alpha^{n-2}\hat{\epsilon}_{u}^{n} + \binom{n-1}{1}\alpha^{n-3}\hat{\epsilon}_{u}^{n-1}\epsilon_{d}+\dots + \binom{n-1}{n-2}\hat{\epsilon}_{u}^{2}\epsilon_{d}^{n-2}].
    \end{align} 
    By the induction hypothesis, we know that for all $k<n$:
    \begin{equation*}
        \E[\hat{\epsilon}_{u}^{k}] = (k-1)\E[\hat{\epsilon}_{u}^{k-2}]\E[\hat{\epsilon}_{u}^{2}].
    \end{equation*}
    Then,
    \begin{align*}
        & \binom{n-1}{k}\alpha^{n-k-2}\E[\hat{\epsilon}_{u}^{n-k}]\E[\epsilon_{d}^{k}] =
        \binom{n-1}{k}\alpha^{n-k-2}\E[\hat{\epsilon}_{u}^{n-k-2}]\E[\hat{\epsilon}_{u}^{2}]\E[\epsilon_{d}^{k}](n-k-1).
    \end{align*}
    Note that:
    \begin{align*}
        & \binom{n-1}{k}(n-k-1) = \frac{(n-1)!(n-k-1)}{(k)!(n-k-1)!}
        = \frac{(n-2)!(n-1)}{(k)!(n-k-2)!} = \binom{n-2}{n-k-2}(n-1).
    \end{align*}
    Therefore,
    \begin{equation}\label{eq: substitution n-1->n-2}
    \begin{gathered}
        \binom{n-1}{k}\alpha^{n-k-2}\E[\hat{\epsilon}_{u}^{n-k}]\E[\epsilon_{d}^{k}] = 
        (n-1)\E[\hat{\epsilon}_{u}^{2}]\binom{n-2}{n-k-2}\E[\hat{\epsilon}_{u}^{n-k-2}]\E[\epsilon_{d}^{k}].
    \end{gathered}
    \end{equation}
    Substituting all the terms except the first one in \eqref{eq: D^(n-1)Z} using  \eqref{eq: substitution n-1->n-2}, we have:
    \begin{equation*}
    \begin{split}
        &\E[D^{n-1}Z] =\E[\alpha^{n-2}\hat{\epsilon}_{u}^{n}] 
        +(n-1)\E[\hat{\epsilon}_{u}^2]\E\left[\sum_{k=0}^{n-2}\binom{n-2}{k}(\alpha\hat{\epsilon}_u)^{k}\epsilon_d^{n-2-k}\right] \\
        & -(n-1)\alpha^{n-2}\E[\hat{\epsilon}_{u}^{n-2}]\E[\hat{\epsilon}_{u}^2] 
        =\alpha^{n-2}\E[\hat{\epsilon}_{u}^{n}] + (n-1) \E[\hat{\epsilon}_u^2]\E[D^{n-2}] - (n-1)\alpha^{n-2}\E[\hat{\epsilon}_{u}^{n-2}]\E[\hat{\epsilon}_{u}^2]
    \end{split}
    \end{equation*}
    Consequently,
    \begin{equation}\label{eq_app: D^(n-1)Z - (n-1)U^2Z^(n-2)}
    \begin{gathered}
        \E[D^{n-1}Z] - (n-1)\E[\hat{\epsilon}_{u}^2]\E[D^{n-2}] =
        \alpha^{n-2}\left(\E[\hat{\epsilon}_{u}^{n}] - (n-1)\E[\hat{\epsilon}_{u}^{n-2}]\E[\hat{\epsilon}_{u}^{2}]\right).
    \end{gathered}
    \end{equation}
    Similarly, we can get
    \begin{equation}\label{eq_app: Z^(n-1)D - (n-1)U^2D^(n-2)}
    \begin{gathered}
        \E[Z^{n-1}D] - (n-1)\E[\hat{\epsilon}_{u}^2]\E[Z^{n-2}] =
        \frac{1}{\alpha^{n-2}}\left(\E[\hat{\epsilon}_{u}^{n}] - (n-1)\E[\hat{\epsilon}_{u}^{n-2}]\E[\hat{\epsilon}_{u}^{2}]\right).
    \end{gathered}
    \end{equation}
    Note that the right hand sides of  \eqref{eq_app: D^(n-1)Z - (n-1)U^2Z^(n-2)} and \eqref{eq_app: Z^(n-1)D - (n-1)U^2D^(n-2)} should be equal to zero. Otherwise one can divide \eqref{eq_app: D^(n-1)Z - (n-1)U^2Z^(n-2)} by \eqref{eq_app: Z^(n-1)D - (n-1)U^2D^(n-2)} and get the value of $\alpha^{2n-4}$. This is because  we can obtain $\E[Z^{n-1}D]$ and $\E[D^{n-2}]$ from the observational distribution. The other term in the expression,  $\E[\hat{\epsilon}^2]$, can also be computed from the observation distribution as it equals $\E[DZ]$. To see this, note that 
    \begin{equation*}
        \E[DZ] = \E\left[\hat{\epsilon}_{u}^2 + \hat{\epsilon}_u(\alpha\epsilon_z + \frac{1}{\alpha}\epsilon_d) + \epsilon_d\epsilon_z\right] = \E\left[\hat{\epsilon}_u^2\right].
    \end{equation*}
    Therefore, we can identify $\alpha^2$ uniquely up to its sign since $\alpha^2$ is a real-valued number. Furthermore the sign of $\frac{\alpha_d}{\alpha_z}$ is the same as the sign of the $\alpha_z\alpha_d\E[\epsilon_u^2]$ which is equal to the $\E[DZ]$. Thus, $\frac{\alpha_d}{\alpha_z}$ will be determined uniquely if \eqref{eq: prop norm dist} is not satisfied for $n$ and the proof is complete.
\end{proof}

\begin{customcorr}{\ref{cor:main}}
Under Assumptions \ref{assum: Var(eps_d)>0}, \ref{assum: finite moments} and \ref{assum: eq iff gaussian}, the causal effect $\beta$ can be recovered uniquely if $\epsilon_u$ is non-Gaussian.
\end{customcorr}

\begin{proof}
Based on \cite{brown1983p}[Chapter 30, Theorem 30.1], under Assumption \ref{assum: eq iff gaussian}, 
the condition $\E\left[\hat{\epsilon}_u^{n}\right] = (n-1)\E\left[\hat{\epsilon}_{u}^{n-2}\right]\E\left[\hat{\epsilon}_{u}^2\right]$
is satisfied for any $n\in \mathbb{N}$ if and only if $\hat{\epsilon}_u$ is Gaussian. Thus, based on Theorem \ref{th:main}, the causal effect $\beta$ is identified if $\epsilon_u$ is non-Gaussian. 
\end{proof}

\begin{customthm}{\ref{th: impossibility result}}
   Suppose that the observed variables in linear SCM in \eqref{eq: main SEM} have jointly Gaussian distribution.
    Under Assumptions \ref{assum: Var(eps_d)>0}, \ref{assum: finite moments} and \ref{assum: Var(eps_z)>0}, the total causal effect $\beta$ cannot be identified uniquely.
\end{customthm}
\begin{proof}
    Without loss of generality, we assume that $Z$, $D$ and $Y$ have zero mean and are generated by a model $\M_1$ as follows:
    \begin{equation*}
    \begin{aligned}
        & \M_1: \\
        & U = \epsilon_u,\\
        & Z = \alpha_z U + \epsilon_z,\\
        & D = \alpha_d U + \epsilon_d,\\
        & Y = \beta D + \gamma U + \epsilon_y,
    \end{aligned}
    \end{equation*}
    and $\alpha_d=1$. Otherwise, instead of $U=\epsilon_u$, one can write $U=\alpha_d\epsilon_u$ and rescale other coefficients respectively. 
    Further, we construct a model $\M_2$ as follows:
    \begin{equation*}
    \begin{aligned}
        & \M_2:\\
        & U = \epsilon_u,\\
        & Z = \frac{1}{k} \alpha_z U + \epsilon_z',\\
        & D =  k U + \epsilon_d',\\
        & Y = \beta' D + \gamma' U + \epsilon_y',
    \end{aligned}
    \end{equation*}
     where $\beta\neq \beta'$ and all the exogenous noises are Gaussian with a mean equal to 0 such that:
    \begin{equation} \label{eq: proof neccessary and sufficient condition}   
    \begin{gathered}
        \Var{Z}^{\M_1} = \Var{Z}^{\M_1}, \quad \Var{D}^{\M_1}=\Var{D}^{\M_2}, \quad \Var{Y}^{\M_1} = \Var{Y}^{\M_2},\\
        \Cov{Z}{D}^{\M_1} = \Cov{Z}{D}^{\M_2}, \quad \Cov{Z}{Y}^{\M_1} = \Cov{Z}{Y}^{\M_2}, \quad \Cov{D}{Y}^{\M_1} = \Cov{D}{Y}^{\M_2}.
    \end{gathered}
    \end{equation}
    Since in both cases $Z$, $D$ and $Y$ are jointly Gaussian then both model agree on the distribution of observed variables.
    The latter means that total causal effect $\beta$ is not identifiable since $\beta\neq \beta'$ and it is impossible to distinguish between them having only observations of $Z$, $D$ and $Y$.

    More specifically, we define $k = 1-\delta$, where $\delta$ some real number such that:
    \begin{align}
        \label{eq: delta inequality}
        & 0 <  \delta < 1 - \sqrt{\frac{\alpha_z^2\Var{\epsilon_u}}{\alpha_z^2\Var{\epsilon_u} + \Var{\epsilon_z}}},\\
        \label{eq: k inequality}
        & \frac{\Var{\epsilon_d}}{\Var{\epsilon_u}} \geq (1-k^2).
    \end{align}
     Accordingly, we define random variables $\epsilon_z'$, $\epsilon_d'$, $\epsilon_y'$ as Gaussian random variables with mean zero having the variances as follows:
    \begin{align*}
        & \Var{\epsilon_z'} := \sigma_z' = \alpha_z^2\Var{\epsilon_u} + \Var{\epsilon_z} - \frac{1}{k^2}\alpha_z^2\Var{\epsilon_u}\\
        & \Var{\epsilon_d'} := \sigma_d' = \Var{\epsilon_u} + \Var{\epsilon_d} - k^2\Var{\epsilon_u} > \Var{\epsilon_d}\\
        & \Var{\epsilon_y'} := \sigma_y' = (\beta + \gamma)^2\Var{\epsilon_u} + \beta^2\Var{\epsilon_d} + \Var{\epsilon_y} - (k\beta' + \gamma')^2\Var{\epsilon_u} - \beta'^2\sigma_d',
    \end{align*}
     where
     \begin{align*}
         & \beta' := \beta + \gamma\Var{\epsilon_u}\left( \frac{1-k^2}{\sigma_{d}'} \right),\\
         & \gamma' := k \gamma \frac{\Var{\epsilon_d}}{\sigma_d'}.
     \end{align*}
     Further we will show that $\sigma_z'>0$, $\sigma_y'>0$ and such that the conditions in \eqref{eq: proof neccessary and sufficient condition} hold, which completes the proof.
     \paragraph{1.}\textit{Here we will prove that} $\sigma_z'>0$ \textit{and}
     \begin{align*}
         & \Var{Z}^{\M_1} = \Var{Z}^{\M_2}, \quad \Cov{Z}{D}^{\M_1} = \Cov{Z}{D}^{\M_2}.
     \end{align*}
     From the inequality \eqref{eq: delta inequality}, we have:
     \begin{align*}
         & k > \sqrt{\frac{\alpha_z^2\Var{\epsilon_u}}{\alpha_z^2\Var{\epsilon_u} + \Var{\epsilon_z}}} \Longrightarrow k^2 > \frac{\alpha_z^2\Var{\epsilon_u}}{\alpha_z^2\Var{\epsilon_u} + \Var{\epsilon_z}} \Longrightarrow \\
         & \alpha_z^2\Var{\epsilon_u} + \Var{\epsilon_z} > \frac{\alpha_z^2}{k^2}\Var{\epsilon_u} \Longrightarrow \sigma_z'= \alpha_z^2\Var{\epsilon_u} + \Var{\epsilon_z} - \frac{\alpha_z^2}{k^2}\Var{\epsilon_u} > 0.
     \end{align*}
     By the definition,
     \begin{align*}
          \Var{Z}^{\M_2} &= \frac{\alpha_z^2}{k^2}\Var{\epsilon_u} + \Var{\epsilon_z'} = \frac{\alpha_z^2}{k^2}\Var{\epsilon_u} + \alpha_z^2\Var{\epsilon_u} + \Var{\epsilon_z} - \frac{\alpha_z^2}{k^2}\Var{\epsilon_u} = \\
          &\alpha_z^2\Var{\epsilon_u} + \Var{\epsilon_z}  = \Var{Z}^{\M_1}, 
     \end{align*}
     and 
     \begin{equation*}
         \Cov{Z}{D}^{\M_1} = \alpha_z\Var{\epsilon_u} =  \Cov{Z}{D}^{\M_2}.
     \end{equation*}

     \paragraph{2.}\textit{ Here we will prove that} $\Var{D}^{\M_1} = \Var{D}^{\M_2}$.

     By the definition,
     \begin{equation}
     \label{eq: var(D)^M_1=var(D)^M_2}
     \begin{split}
         \Var{D}^{\M_2} =& k^2 \Var{\epsilon_u} + \Var{\epsilon_d'} = k^2 \Var{\epsilon_u} + \Var{\epsilon_u} + \Var{\epsilon_d} - k^2\Var{\epsilon_u} = \\
         & \Var{\epsilon_u} + \Var{\epsilon_d} = \Var{D}^{\M_1}.
     \end{split}
     \end{equation}

     \paragraph{3.} \textit{Here we will prove that:}
     \begin{align*}
         \Cov{Z}{Y}^{\M_1} = \Cov{Z}{Y}^{\M_2}, \quad \Cov{D}{Y}^{\M_1}=\Cov{D}{Y}^{\M_2}.
     \end{align*}

     By the definition,
     \begin{align*}
         \Cov{Z}{Y}^{\M_2} &= \frac{1}{k}\alpha_z(\beta' k + \gamma')\Var{\epsilon_u} = \alpha_z\left(  \beta + \gamma\Var{\epsilon_u}\left( \frac{1-k^2}{\sigma_{d}'} \right) + \gamma \frac{\Var{\epsilon_d}}{\sigma_d'}\right)\Var{\epsilon_u} = \\
         & \alpha_z \left( \beta + \gamma \frac{\Var{\epsilon_u} + \Var{\epsilon_d} - k^2\Var{\epsilon_u}}{\sigma_d'} \right)\Var{\epsilon_u} = 
         \alpha_z(\beta + \gamma)\Var{\epsilon_u} = \Cov{Z}{Y}^{\M_1}
     \end{align*}
     and 
     \begin{align*}
        \Cov{D}{Y}^{\M_2} &=  \beta'\Var{D} + k \gamma'\Var{\epsilon_{u}} = \left( \beta + \gamma\Var{\epsilon_u}\left( \frac{1-k^2}{\sigma_{d}'} \right) \right) \Var{D} + k^2 \gamma \frac{\Var{\epsilon_d}}{\sigma_d'} \Var{\epsilon_u} = \\
        & \beta \Var{D} + \gamma\Var{\epsilon_u}\left( \frac{1-k^2}{\sigma_{d}'} \right) \left( \Var{\epsilon_u} + \Var{\epsilon_d} \right) + k^2 \gamma \frac{\Var{\epsilon_d}}{\sigma_d'} \Var{\epsilon_u} = \\
        & \beta \Var{D} + \gamma \Var{\epsilon_u} \frac{(1-k^2)\Var{\epsilon_u} + \Var{\epsilon_d}}{\sigma_d'} = \beta\Var{D} + \gamma\Var{\epsilon_u} = \Cov{D}{Y}^{\M_1}
     \end{align*}
     \paragraph{4.} \textit{Here we will proof that} $\sigma'_y \geq 0$  \textit{and} $\Var{Y}^{\M_1}=\Var{Y}^{\M_2}$.
     
     To get inequality $\sigma'_y\geq0$ it is enough to show that
    \begin{equation*}
        (\beta+\gamma)^2\Var{\epsilon_u} + \beta^2\Var{\epsilon_d} \geq  (k\beta' + \gamma')^2\Var{\epsilon_u} + \beta'^2\sigma_d'.
    \end{equation*}
    Therefore
    \begin{align*}
        & (\beta+\gamma)^2\Var{\epsilon_u} + \beta^2\Var{\epsilon_d} \geq  (k\beta' + \gamma')^2\Var{\epsilon_u} + \beta'^2\sigma_d' \iff\\
        & (\beta+\gamma)^2\Var{\epsilon_u} + \beta^2\Var{\epsilon_d} \geq k^2\left( \beta + \gamma\Var{\epsilon_u}\left( \frac{1-k^2}{\sigma_{d}'} \right) + \gamma \frac{\Var{\epsilon_d}}{\sigma_d'} \right)^2 \Var{\epsilon_u} + \beta'^2\sigma_d' \iff \\
        & (\beta+\gamma)^2\Var{\epsilon_u} + \beta^2\Var{\epsilon_d} \geq k^2(\beta+\gamma)^2 + \left( \beta + \gamma\Var{\epsilon_u}\left( \frac{1-k^2}{\sigma_{d}'} \right) \right)^2 \sigma_d' \iff \\
        & (1-k^2)(\beta+\gamma)^2\Var{\epsilon_u} + \beta^2\Var{\epsilon_d} \geq \beta^2\Var{\epsilon_d'} + 2\beta\gamma\Var{\epsilon_u}(1-k^2) + (1-k^2)^2C,
    \end{align*}
    where $C = \gamma^2\frac{\Var{\epsilon_u}^2}{\sigma_d'}$. From  \eqref{eq: var(D)^M_1=var(D)^M_2}, we can get
    \begin{align*}
         & (1-k^2)(\beta+\gamma)^2\Var{\epsilon_u} + \beta^2\Var{\epsilon_d} \geq \beta^2\Var{\epsilon_d'} + 2\beta\gamma\Var{\epsilon_u}(1-k^2) + (1-k^2)^2C \iff \\
         & (1-k^2)(\beta+\gamma)^2\Var{\epsilon_u} \geq (1-k^2)\beta^2\Var{\epsilon_u} + 2\beta\gamma\Var{\epsilon_u}(1-k^2) + (1-k^2)^2C \iff\\
         & \gamma^2\Var{\epsilon_u} \geq (1-k^2)C.
    \end{align*}
    From the inequality \eqref{eq: k inequality} we have
    \begin{align*}
         \gamma^2\Var{\epsilon_u} \geq \gamma^2(1-k^2)\frac{\Var{\epsilon_u}^2}{\Var{\epsilon_d}} \geq \gamma^2(1-k^2)\frac{\Var{\epsilon_u}^2}{\sigma_d'} = (1-k^2)C.
    \end{align*}
    The last inequality follows from the fact that $\sigma_d' \geq \Var{\epsilon_d}$ (the definition of $\sigma_d'$). Therefore in our construction for the second model, we have $\sigma'_y \geq 0$.

    Finally, 
    \begin{align*}
        \Var{Y}^{\M_2} &= (\beta'k + \gamma')^2\Var{\epsilon_u} + \beta'^2\sigma_d' + \Var{\epsilon_y'} = \\
        & (\beta'k + \gamma')^2\Var{\epsilon_u} + \beta'^2\sigma_d' + \\
        & (\beta + \gamma)^2\Var{\epsilon_u} + \beta^2\Var{\epsilon_d} + \Var{\epsilon_y} - (k\beta' + \gamma')^2\Var{\epsilon_u} - \beta'^2\sigma_d' =\\
        & (\beta + \gamma)^2\Var{\epsilon_u} + \beta^2\Var{\epsilon_d} + \Var{\epsilon_y} = \Var{Y}^{\M_1}.
    \end{align*}

    The above claims show that the two models are indistinguishable from the observational distribution and they have different causal effects of $D$ on $Y$ and thus the proof is complete.
\end{proof}

\newpage
\section{Experiments}
\label{app: experiments}
\subsection{Synthetic data}

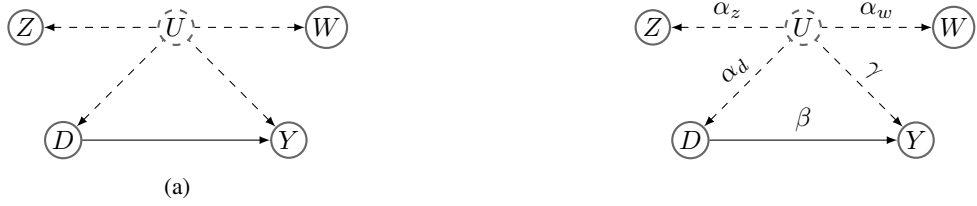
\begin{figure}[t]
    \centering
    \begin{subfigure}[t]{0.4\linewidth}
        \centering
        \begin{tikzpicture}[
            roundnode/.style={circle, draw=black!60,, fill=white, thick, inner sep=1pt},
            dashednode/.style = {circle, draw=black!60, dashed, fill=white, thick, inner sep=1pt},
            ]
            \node[roundnode]        (D)        at (-1.5, 0)                   {$D$};
            \node[roundnode]        (Y)        at (1.5, 0)                    {$Y$};
            \node[roundnode]        (Z)        at (-2, 1.5)                   {$Z$};
            \node[roundnode]        (W)        at (2, 1.5)                    {$W$};
            \node[dashednode]       (U)        at (0, 1.5)                    {$U$};
            
            \draw[-latex] (D.east) -- (Y.west) node[midway,sloped,above] {} ;
            \draw[latex-, dashed] (D) -- (U) node[midway,sloped,above] {};
            \draw[latex-, dashed] (Y) -- (U) node[midway,sloped,above] {};
            \draw[latex-, dashed] (Z) -- (U) node[midway,sloped,above] {};
            \draw[latex-, dashed] (W) -- (U) node[midway,sloped,above] {};
            
        \end{tikzpicture}
        \caption{}
        \label{fig: DiD graph_app}
    \end{subfigure}
    \hfill
    \begin{subfigure}[t]{0.4\linewidth}
        \centering
        \begin{tikzpicture}[
            roundnode/.style={circle, draw=black!60,, fill=white, thick, inner sep=1pt},
            dashednode/.style = {circle, draw=black!60, dashed, fill=white, thick, inner sep=1pt},
            ]
            \node[roundnode]        (D)        at (-1.5, 0)                   {$D$};
            \node[roundnode]        (Y)        at (1.5, 0)                    {$Y$};
            \node[roundnode]        (Z)        at (-2, 1.5)                   {$Z$};
            \node[roundnode]        (W)        at (2, 1.5)                    {$W$};
            \node[dashednode]       (U)        at (0, 1.5)                    {$U$};
            
            \draw[-latex] (D.east) -- (Y.west) node[midway,sloped,above] {$\beta$} ;
            \draw[latex-, dashed] (D) -- (U) node[midway,sloped,above] {$\alpha_d$};
            \draw[latex-, dashed] (Y) -- (U) node[midway,sloped,above] {$\gamma$};
            \draw[latex-, dashed] (Z) -- (U) node[midway,sloped,above] {$\alpha_z$};
            \draw[latex-, dashed] (W) -- (U) node[midway,sloped,above] {$\alpha_w$};
        \end{tikzpicture}
        \label{fig: main graph_app}
    \end{subfigure}
    \caption{ The considered causal graph for the experiments with linear assignments in the SCM framework.}
\end{figure}
For the experiments with synthetic data we assume that the samples are generated according to the following linear SCM:
\begin{equation}
\label{eq_app: SEM for exp}
\begin{split}
    U &:= \epsilon_u, \\
    W &:= \alpha_w U + \epsilon_w, \\
    Z &:= \alpha_z U + \epsilon_z, \\
    D &:= \alpha_d U + \epsilon_d, \\
    Y &:= \beta D + \gamma U +\epsilon_y.
\end{split}
\end{equation}
Given the observational data, we estimate the value of $\beta$ using the following methods and report the performances against the number of observed samples.
\begin{enumerate}
    \item Cross-Moment algorithm proposed in this work.
    \item DiD method according to \eqref{eq: TWFE regression}.
    \item A simple linear regression model based on the following equation:
        $\hat{Y} = \alpha Z + \hat{\beta} D.$
    \item The ``two-proxy'' method in \cite{kuroki2014measurement}. 
\end{enumerate}
For each value of sample size, we ran an experiment $10$ times and reported the 
average relative error:
\begin{equation*}
    err = \E\left[\left|\frac{\beta-\hat{\beta}}{\beta}\right|\right],
\end{equation*}
and the standard deviation with the colored regions on the plots.
Before each run, we randomly generate parameters $\alpha_d$, $\alpha_z$, $\alpha_w$, $\beta$, $\gamma$ as follows:
\begin{itemize}
    \item $\alpha_d$ is randomly sampled from the interval $(-2, -0.2) \cup (0.2, 2)$,
    \item $\alpha_z$, $\beta$, $\gamma$ are randomly sampled such that the absolute value of the ratios between $\alpha_d$ and each of the variables $\alpha_z$, $\beta$, $\gamma$ are in the interval $(0.2, 2)$,
    \item We set $\alpha_w=\alpha_z$ to have a consistent setting for measuring the effect of noise in proxy variables on the Cross-Moment method and the two-proxy method proposed by \cite{kuroki2014measurement}.
\end{itemize}

In our experiments, the variances of $\epsilon_z$, $\epsilon_d$, $\epsilon_y$ are 10 time less than the variance of $\epsilon_u$. We also set the variance of $\epsilon_w$ to be 10 time bigger than the variance of $\epsilon_u$. Thus, proxy $W$ is much noisier compared with $Z$.   

In the case of having two proxy variables $W, Z$, we combine the results of the Cross-Moment method applied for each proxy separately and called these overall procedure as ``Cross-Moment: $W-Z$'' method. More precisely ''Cross-Moment: $W-Z$`` method works as follows:
\begin{enumerate}
    \item For $i$ in $[1:t]$, randomly sample with replacement some portion of all observational data $(\mathbf{Z}, \mathbf{W}, \mathbf{D}, \mathbf{Y})$ that we denote by $(\Z_i, \W_i, \D_i, \Y_i)$.
    \item Using Cross-Moment method over proxy variables $W$ and $Z$ separately, we estimate the causal effect $\beta$ from the data  $(\Z_i, \W_i, \D_i, \Y_i)$. Let us denote this estimates as $\beta^{(i)}_{W}$ and $\beta^{(i)}_{Z}$ accordingly.
    \item Having $\{\beta^{(i)}_{W}\}_{i=1}^{t}$ and $\{\beta^{(i)}_{Z}\}_{i=1}^{t}$,we approximate the variances of the estimates made from proxies $Z$ and $W$. We denote $\sigma_{\beta_W}^2$ and $\sigma_{\beta_Z}^2$, respectively
    \item Compute the final estimation of $\beta$ as follows:
    \begin{equation*}
        \hat{\beta} := \frac{\sigma_{\beta_W}^{-2}}{\sigma_{\beta_W}^{-2} + \sigma_{\beta_Z}^{-2}}\frac{\sum_{i=1}^{t} \beta^{(i)}_{W}}{t} + \frac{\sigma_{\beta_Z}^{-2}}{\sigma_{\beta_W}^{-2} + \sigma_{\beta_Z}^{-2}}\frac{\sum_{i=1}^{t} \beta^{(i)}_{Z}}{t}.
    \end{equation*}
\end{enumerate}

 All the experiments were performed using 16 GB RAM and 12th Gen Intel(R) Core(TM) i7-12700H   2.30 GHz.


\subsubsection{Exponential distribution}
Here we assume that all exogenous noises $\epsilon_u$, $\epsilon_z$, $\epsilon_w$, $\epsilon_d$, $\epsilon_y$ are from the class of exponential distributions. At the beginning of each run, we randomly choose the variance for the exogenous noise $\epsilon_u$ from the interval $(1, 10)$ and set  all other exogenous noise distributions as we discussed in the previous section.   

In addition to the experiments presented in the Section \ref{sef: Synthetic data} on Exponetial distributions we also illustrate the performance of methods against the ratio between $\Var{\epsilon_{w}}$ to $\Var{\epsilon_{u}}$ in Figure \ref{fig_app: exponential perf&rate}. To sum up we observe that two-proxy method suffers much more from the noise in the proxies and is much less stable than our ``Cross-Moment W-Z'' method.

\subsubsection{Uniform distribution}
Here we assume that all the exogenous noises $\epsilon_u$, $\epsilon_z$, $\epsilon_w$, $\epsilon_d$, $\epsilon_y$ have uniform distribution. In this scenario, we considered the same setting as for the exponential distributions. At the beginning of each run, we consider the exogenous noise $\epsilon_u$ to be a uniform distribution on the interval $[-a, a]$, where $a$ is a random real number picked from the interval $(1, 10)$.

\begin{figure}
\begin{subfigure}[t]{0.45\textwidth}
        \includegraphics[width=7cm]{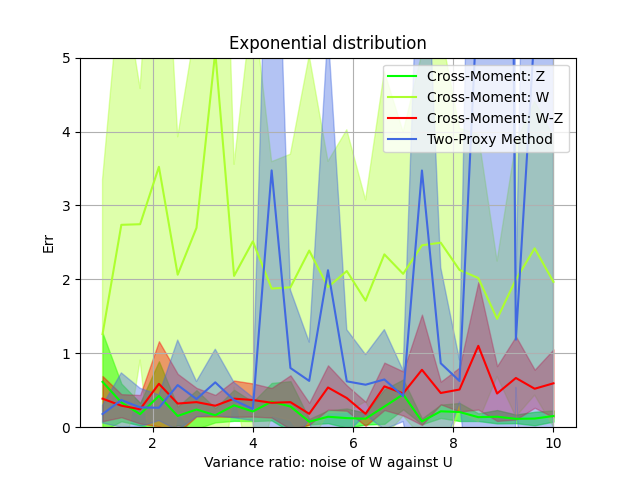}
        \caption{}
        \label{fig_app: exponential perf&rate}
    \end{subfigure}
    \hfill
    \begin{subfigure}[t]{0.45\textwidth}
        \includegraphics[width=7cm]{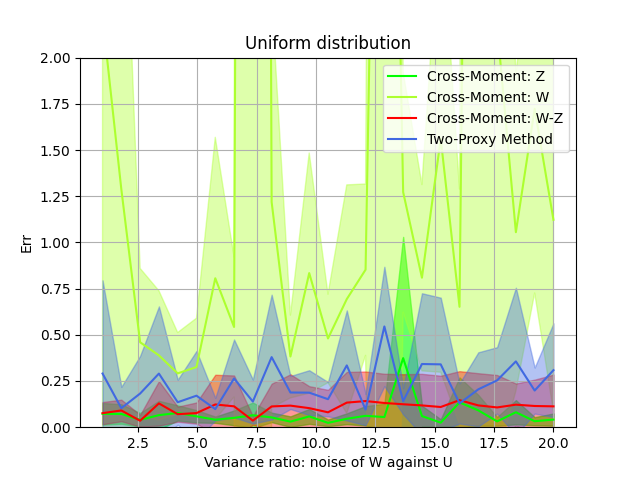}
        \caption{}
        \label{fig_app: uniform perf&rate}
    \end{subfigure}
    \caption{The performance measure $err$ against $\Var{\epsilon_w}/\Var{\epsilon_u}$.}
\end{figure}

\begin{figure}
    \centering
    \includegraphics[width=0.45\textwidth]{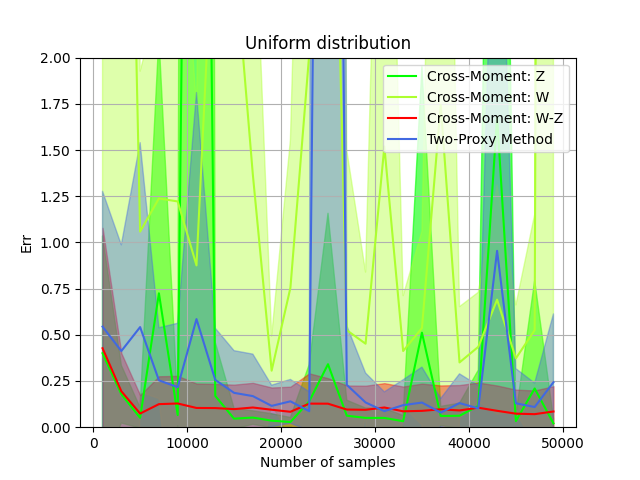}
    \caption{The performance measure $err$ against the number of samples. Colored regions represent the standard deviation of $err$.}
    \label{fig_app: uniform perf&sample size}
\end{figure}

 Figure \ref{fig_app: uniform perf&sample size}  illustrates  the performance of the Cross-Moment and Two-Proxy methods with respect to the number of observed samples. Again, we observe that ``Cross-Moment W-Z'' method more stable than Two-Proxy method. Additionally, in Figure \ref{fig_app: uniform perf&rate}, we show the dependence of the performance of the methods on the ratio between $\Var{\epsilon_{w}}/\Var{\epsilon_{u}}$. Although, for the uniform distribution two-proxy method is more stable but  ``Cross-Moment W-Z'' algorithm still performs better for high values of $\Var{\epsilon_{w}}/\Var{\epsilon_{u}}$.

\newpage
\section{Derivations of DiD Estimator in Linear SCMs}
\label{app: DiD}
Without loss of generality, we assumed that all the variables in the system are mean zero. Thus, there is no intercept term in the linear regression model: $\hat{Y}= \hat{\beta}_1 T + \hat{\beta}_2D +\hat{\beta} DT$.
Let $z_i$ be the outcome of the individual $i$ before assigning treatment. The mean of squared residual over the population before treatment is $\sum_{i}(z_i-\hat{\beta}_2d_i)/n$ where $n$ is the size of the population and $d_i\in \{0,1\}$ is equal to one if the treatment is assigned to individual $i$. Otherwise, $d_i$ is zero. For the post-treatment phase, let $y_i(d_i)$ be the outcome of individual $i$. Hence, the mean of squared residual over the population in the post-treatment phase is: $\sum_i (y_i(d_i)-\hat{\beta}_1-(\hat{\beta}_2+\hat{\beta})d_i)^2/n$. By performing a linear regression on the whole samples observed in pre-treatment and post-treatment phases over the population, we are minimizing the following risk: $\sum_{i}(z_i-\hat{\beta}_2d_i)/n +\sum_i (y_i(d_i)-\hat{\beta}_1-(\hat{\beta}_2+\hat{\beta})d_i)^2/n$. Considering the uniform distribution among the individuals, the objective function in the minimization is equivalent to: $\min_{\hat{\beta}_1, \hat{\beta}_2, \hat{\beta}}\E[(Z-\hat{\beta}_2D)^2]+\E[(Y-\hat{\beta}_1-(\hat{\beta}+\hat{\beta}_2)D)^2]$. By taking partial derivative with respect to $\hat{\beta}_1$, $\hat{\beta}_2$, and $\hat{\beta}$ and setting them  to zero, we can imply that $\hat{\beta}_1=0$, $\hat{\beta}_2=\E[ZD]/\E[D^2]$, and $\hat{\beta}=(\E[YD]-\E[ZD])/\E[D^2]$, respectively. According to linear SCM in \eqref{eq: main SEM}, we have $\E[YD]=\alpha_d(\alpha_d\beta+\gamma)\Var{\epsilon_u}+ \beta \Var{\epsilon_d}$, $\E[ZD]=\alpha_d\alpha_z\Var{\epsilon_u}$, and $\E[D^2]=\alpha_d^2\Var{\epsilon_u}+\Var{\epsilon_d}$. By plugging these terms in the equation for $\hat{\beta}$, we get the equation in \eqref{eq:DiD linear SCM}.

\end{document}